\documentclass{aastex62}
\usepackage[utf8]{inputenc}
\usepackage{graphicx}

\submitjournal{PASP}

\shorttitle{The LRIS Longslit Pipeline}
\shortauthors{Perley}

\begin{document}

\title{Fully-Automated Reduction of Longslit Spectroscopy with the \\ Low Resolution Imaging Spectrometer at Keck Observatory}
\email{d.a.perley@ljmu.ac.uk}

\author[0000-0001-8472-1996]{Daniel A. Perley}
\affil{Astrophysics Research Institute, Liverpool John Moores University, \\ IC2, Liverpool Science Park, 146 Brownlow Hill, Liverpool L3 5RF, UK}

\begin{abstract}
I present and summarize a software package (``LPipe'') for completely automated, end-to-end reduction of both bright and faint sources with the Low-Resolution Imaging Spectrometer (LRIS) at Keck Observatory.  It supports all gratings, grisms, and dichroics, and also reduces imaging observations, although it does not include multislit or polarimetric reduction capabilities at present.  It is suitable for on-the-fly quicklook reductions at the telescope, for large-scale reductions of archival data-sets, and (in many cases) for science-quality post-run reductions of PI data.  To demonstrate its capabilities the pipeline is run in fully-automated mode on all LRIS longslit data in the Keck Observatory Archive acquired during the 12-month period between August 2016 and July 2017.  The reduced spectra (of 675 single-object targets, totaling $\sim$200 hours of on-source integration time in each camera), and the pipeline itself, are made publicly available to the community.
\end{abstract}

\section{Introduction} \label{sec:intro}
Slit spectroscopy provides the cheapest and simplest way of obtaining spectral information at more than one position on the sky.  Most classical optical spectrographs support some sort of longslit mode, and many instruments also support the use of customized slit-mask plates to permit the observation of multiple objects that are not spatially collinear.

Most commonly, the desired output of a slit-based observation is a flux-calibrated, one-dimensional spectrum ($\lambda$ vs $F_\lambda$) of an object or a group of objects.\footnote{In other cases, such as in measurement of galaxy rotation curves, a two-dimensional map of position versus spectrum is desired.}
Production of science-quality data requires many reduction steps, including: (1) bias correction and flat-fielding, (2) removal of CCD artifacts and cosmic rays, (3) identification of the source/sources of interest on the slit, modeling the traces their dispersed light makes across the detector, and extraction of these traces,  (4) wavelength calibration using arc lamps and/or night sky lines, (5) flux calibration, including correcting for effects arising from Earth's atmosphere such as telluric absorption.   Additionally, if multiple exposures are taken of the same target these must be combined together.  If multiple instrument setups are used to cover a wider wavelength region, these must be similarly combined.  The impacts of order cross-contamination, scattered/reflected light, and instrument flexure must also be dealt with or at least understood.   

As a result, production of even a single science-quality spectrum can take a trained observer many hours (possibly, many days) of effort.  This is particularly true when operating a new or unfamiliar instrument.  These delays push back the timescale for discovery and publication of interesting results.  This problem can be particularly acute when observing transient objects or conducting target-of-opportunity observations: in these cases, rapid characterization of the source (redshift measurement of a GRB afterglow or classification of a young supernova) is often critical to planning further follow-up.  Assistance with these reductions may not be available during the middle of the night when such observations are necessarily obtained.

For these reasons, spectroscopic pipelines are becoming commonplace at a variety of observatories.  Many of these pipelines are more accurately described as high-level toolkits, as significant user input and technical knowledge is still required to transform raw data to science-quality spectra.  Other pipelines are nearly automated, but only function in a narrow range of circumstances---suitable for survey projects and for telescopes whose configurations are inflexible, but of limited applicability to the wider user community for more general-purpose instruments.   Also, although it has recently become commonplace to require pipeline or reduction capabilities for newly-developed spectrographs as part of their construction costs, it is difficult for observatories to motivate (fund) the production of similar official pipelines for complex legacy instruments, even when they are still in common use.

One such well-established spectrograph is the Low Resolution Imaging Spectrometer (LRIS) at Keck Observatory, the first science-quality faint-object spectrograph on a 10~m-class optical telescope and still among the most commonly-used and productive instruments in astronomy \citep{Kulkarni+2016}.  Thanks to several upgrades over its lifetime LRIS is a a complex and flexible instrument: it employs an atmospheric dispersion corrector, two separate channels (red and blue) with separate detectors, dispersers and filter sets, and has multi-object capabilities.

Despite its long history LRIS currently has no widely-available reduction pipeline.  A few observatory-maintained IRAF packages\footnote{\url{https://www2.keck.hawaii.edu/inst/lris/kecklris.html}} exist for basic tasks.  A high-quality advanced reduction toolset for LRIS and many other single- and dual-channel spectrographs has been developed  (\texttt{LowRedux}\footnote{\url{https://github.com/profxj/xidl}; a modified version of this package optimized for a different spectrograph is described in \cite{Bernstein+2015}.}, by J. X. Prochaska, J. Hennawi, and D. Schlegel), but it is not completely automatic in operation and not extensively documented.  The FLAME pipeline \citep{Belli+2017} supports LRIS and a few other instruments, but requires a large amount of user input, does not perform flux calibration or telluric correction, and does not combine the red and blue arms.  A python reduction package for a variety of Keck (and other) instruments is in development\footnote{\url{https://pypeit.readthedocs.io/}}, but is not yet officially released.

In this paper we summarize a spectroscopic reduction pipeline for LRIS.  It is completely automated: given a directory containing a night's worth of raw FITS data, the pipeline will convert these files to science-quality flux-calibrated spectra of all science targets with no user input whatsoever.  The pipeline is interruptible and resumable, logs its progress and actions, and provides extensive header information and other metadata.  The pipeline is optimized for functionality in automatic mode under a wide range of circumstances, but also contains a number of options to permit user control, including interactive graphical tools for controlling the extraction of source traces or inspecting final output.  Robustness of reductions to a wide range of configurations, conditions, and data-acquisition methods are prioritized over obtaining the highest-possible S/N or calibration precision.

An imaging pipeline for LRIS is also available, although we do not describe it in detail here.  The pipeline does not yet support LRIS's multi-object capabilities or polarimetry, although this is a long-term goal.

The pipeline has been widely used, mainly by the supernova groups at UC Berkeley and Caltech, since it was originally developed in 2013.  In this paper we provide official, formal documentation of its functionality and capabilities, and demonstrate its capabilities by applying it to a large data set from the public LRIS archive.

The pipeline is written in IDL, a proprietary language that (although formerly common in astronomy) unfortunately makes direct reuse of the code for other instruments limited.  The techniques described here may nevertheless be useful for inspiring or guiding the development of similar high-level pipelines for other facilities in more modern languages.

The reader should be aware that as with most software packages, the pipeline is in continuous development.  While the descriptions below are accurate at the time of submission of this paper (and describe version 2019.03 of the pipeline), future changes are likely to modify the behavior of the pipeline to some extent, and users should consult the project website for the most up-to-date description of its capabilities.

\section{Summary of LRIS}

A full technical description of the LRIS instrument can be found in the original paper (\citealt{Oke+1995}, which describes only the original red channel).  The blue channel, added in 2000, is described by \cite{McCarthy+1998} and \cite{Steidel+2004}.  The atmospheric dispersion corrector (generic to Keck I), added in 2007, is described in \cite{Phillips+2006} and spectropolarimetric capabilities---not supported by this pipeline---are described in \cite{Goodrich+1995}.  The upgraded red detector (added in June 2009 and replaced in November 2010) is summarized in \cite{Rockosi+2010}.  We do not attempt to duplicate these discussions, but do provide a brief summary of relevant capabilities for the observer, with emphasis on how LRIS compares to other facilities.

\begin{figure}[tbp] 
   \centering
   \includegraphics[width=4.5in]{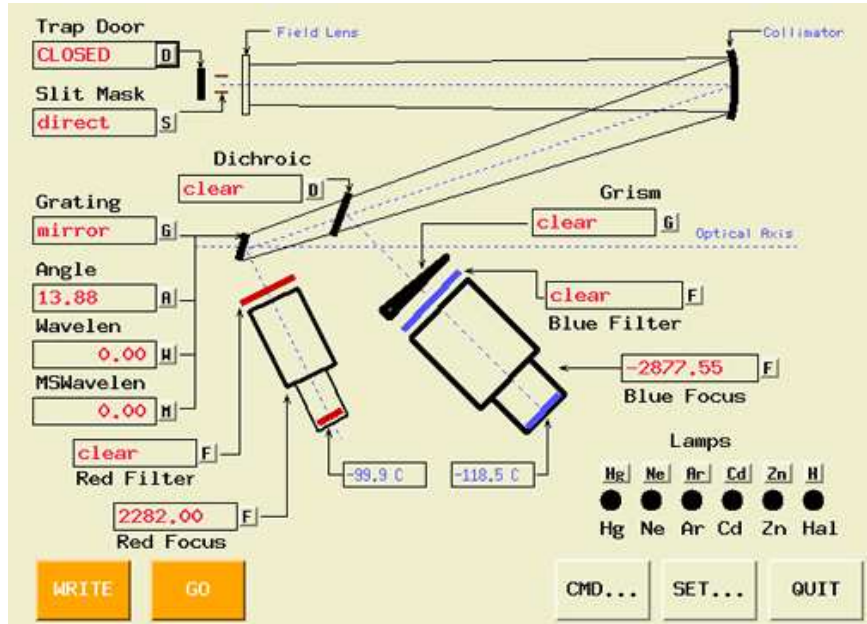}
   \caption{\small
   The XLRIS configuration tool seen by observers at the telescope, which demonstrates the basic optical configuration of the instrument and options available to reconfigure it.
 }
\label{fig:xlris} 
\end{figure}

The basic configuration of LRIS---familiar to anyone who has operated the instrument from the observing GUI in the telescope control room---is illustrated in Figure \ref{fig:xlris}.  The instrument is located at the Cassegrain focus of Keck I.  Light enters through a trapdoor, passes through a slit or slitmask positioned at the focal plane, and continues through collimating optics.  A dichroic normally splits the light into red (transmitted) and blue (reflected) components---the separation wavelength chosen by most observers is 5600\,\AA, but several alternative dichroics can be used, and the dichroic can also be removed (for red operations only) or replaced by a mirror (for blue operations only).  Dispersive elements---a grating on the red side and a grism on the blue side---split the light into component wavelengths.  Several red gratings (providing effective resolutions from 0.4 to 3 \AA\ per resolution element) are available and the grating angle can be changed to control the wavelength range. Several blue grisms are likewise available (resolutions between 1.6 and 9 \AA), although the blue wavelength range is fixed for each grism.  Dispersed light is sent to the detectors: a pair of red-sensitive deep-depletion 2048$\times$4096 CCDs in the red camera and a pair of blue-optimized 2048$\times$4096 CCDs in the blue camera.  Each camera has its own separate shutter (not shown in Figure \ref{fig:xlris}).  Calibration lamps can be turned on to illuminate the back of the trapdoor.

Several of these features are worthy of note for performing pipeline reductions.  First, the presence of two channels (red and blue) enables very broad wavelength coverage, providing continuous observations from the atmospheric UV cutoff to the sensitivity limit of the detector well beyond 10000\,\AA\ in low-resolution mode.  This greatly increases the complexity of observing and reductions---the requirements of the two cameras are often competing (when choosing exposure times, dithering methods, calibrations, and so on), and production of a final ``combined'' spectrum requires putting together two spectra of different resolutions, detector characteristics, and calibration files.  Furthermore, the imposition of a dichroic causes strongly wavelength-dependent attenuation in the response curve in both detectors, directly in the middle of the wavelength range of the combined spectrum.

Second, as nearly every element can be moved or changed, the number of observational types that a comprehensive pipeline must support is large.  There is no single ``standard'' setup.  Furthermore, since Keck is an entirely classically-scheduled facility, observing and calibration procedures vary even for fundamentally similar observational configurations.  Inexperienced users may on occasion commit mistakes in data acquisition that complicate later reduction.

Third, LRIS suffers from significant flexure and has no flexure compensation system.  Observations at different positions and orientations can exhibit wavelength offsets in excess of several \AA.  Corrections to the wavelength solution must be made to avoid significant errors in the wavelength calibration, which in turn would propagate to errors in flux calibration and telluric correction.  Spatial flexure is also significant, and can exacerbate problems associated with irregularities in the slit.

Finally, the current LRIS CCDs on both red and blue sides exhibit a spatial gap in their center.  Thus the spatial coverage along the slit exhibits a gap as well, and the primary target of an observation is centered away from the actual center of the longslit (on what we refer to as the ``right'' CCD, as distinguished from the left CCD).  The two chips for each camera have similar properties but (particularly on the blue) do have different wavelength-dependent quantum efficiencies.

\section{Pipeline Operations}

The LRIS pipeline has been developed to handle all the above features, with the exceptions that it does not currently support slitmask observations or polarimetric capabilities and offers only partial support for reduction of the secondary (``left'') CCD on each side.  A driving requirement is that the pipeline is able to provide hands-off, intervention-free, end-to-end reductions for standard observing cases (specifically, for observations of single point sources in uncrowded fields with detectable continuum flux).

In this section we describe, in order, the individual tasks (``steps'') taken by the pipeline to achieve this aim.  In default operations the pipeline automatically executes every task in order until final spectra are produced at the end.  Cautious users desiring more control can also run individual steps or series of steps, or even control reduction steps at the level of individual targets or individual exposures.   We emphasize the spectroscopic pipeline, although a sophisticated imaging reduction pipeline is available as well in the same package; this is briefly summarized in \S \ref{sec:imaging}.  Additional details on pipeline use can be found in the online manual.\footnote{\url{http://www.astro.caltech.edu/~dperley/programs/lris/manual.html}}

\subsection{Data regularization, bias correction, artifact correction}

A variety of tasks are performed effectively simultaneously during the first step of the pipeline (in principle they could be separated, but performing them together saves disk write/reads.)  The data format is simplified and regularized, and the bias level is measured individually for each column using the CCD overscan region and subtracted. 

\emph{Data simplification} - Since 2009, LRIS observations (from both red and blue sides) have been stored in multi-extension FITS files with four extensions, one for each amplifier.  Each of these also contains an overscan region.  These files can occasionally be difficult for users to understand without context: orientation, wavelength direction, etc. are not obvious, there is no WCS information written to the header, and (in the case of the red detector) the order of the four files does not match their physical sequence.  The first step of the pipeline is to join all four amplifiers together into a simple 2D image that can be stored as a single-extension FITS file.

In LRIS raw, multi-extension output frames, the CCD's wavelength axis is vertical ($y$ axis) and the spatial axis is horizontal ($x$ axis).  This convention is not well-suited to display on the wide-screen monitors that are currently in favor (which have far more pixels along the $x$ direction than the $y$ direction) and it also does not match the orientation of the image display as seen in the telescope control room.  To further simplify things for the user, then, the $x$ and $y$ axes are transposed.  This also ensures that for a slit PA of zero, ``up'' on the display is due north and ``down'' is due south.

\textit{Overscan subtraction} - At the same time, the bias level of the CCD is subtracted using the overscan columns.  LRIS exhibits very low read noise and little bias structure beyond what can be removed via the overscan method, so this is usually adequate.  The present version of the pipeline does not yet support construction or subtraction of bias or dark frames, although this capability is planned for the near future.  

\textit{Artifact removal} - A known bad pixel list is loaded and affected pixels are marked with NaN flags.

\textit{Header calculations} - A variety of other parameters are calculated and added to the FITS header.  These include calculation of the mid- and end-times of the exposure, the on-sky slit position angle, the average sky background, information about when the instrument configuration was last change, and the coordinates, elevations, and angular distances of the Sun and Moon.  Interpretive comments are added for most header keywords.

Provisional WCS information (using the standard FITS WCS convention, and recognized by DS9 and other display software) is also added to the header.  For imaging observations this is a standard WCS based on the RA and DEC keywords.  For spectroscopic exposures the (transposed) $x$-axis unit is the \emph{approximate} wavelength in \AA ngstroms while the $y$-axis unit is the distance in arcsec from the central chip gap along the slit.  These are approximate solutions only for the convenience of the user and do not take into account any calibration data nor any nonlinearity.  

The software which saves LRIS images to disk has a curious tendency to drop (not write to disk) certain groups of telescope-related header keywords about 2\% of the time.  As a result, important values such as the target name, position, and slit orientation are often missing (instrument-specific keywords such as the filter or grating angle are almost never affected).  Fortunately because observations are almost always taken in red+blue pairs, the lost information can be recovered by searching for a frame taken simultaneously on the opposite camera (blue or red); the pipeline conducts a check for dropped keywords and replaces them by this procedure.  On rare occasions, both members of the pair have critical keywords dropped, or observations were taken in only one camera at a time (the latter is fairly common during calibrations).  These observations cannot be processed further unless the header keywords are manually readded (the \texttt{editheader} task can be used for this if necessary).

\subsection{Flat-field construction}

It is generally only possible to correct for small-scale (i.e., pixel-to-pixel) response differences on LRIS given the types of calibrations that can feasibly be obtained by typical users.   Although several commonly-used longslits show significant non-uniformities on small scales (\S \ref{sec:flats}), these move on the CCD from field to field due to flexure and cannot be calibrated via a single, global flat-field observation.   Similarly, it is not possible to measure large-scale wavelength-direction illumination differences because these cannot be distinguished from differences in the efficiency of the grating and CCD as a function of wavelength.  As a result, all flat-field construction methods employed by the pipeline average out or otherwise remove most large-scale differences along one or both dimensions, although exactly how this is performed depends on the type of flat-field being processed.  It also means that flats taken with essentially any slit can be used, regardless of the slit actually observed with during the night.

Several types of flat-field are available with LRIS:
\begin{itemize}
\item The simplest type of flat is the internal halogen flat.  The lamp is bright and relatively hot, and observations can be obtained quickly at any position or orientation and at any time, including before the telescope is released or during night-time observations.  However, the halogen lamp produces significant variations in both the intensity of illumination and shape of the spectrum along the slit, and it produces substantial scattered red light into the blue camera that dominates the signal over the NUV range (Figure \ref{fig:flatfields}).

\item Dome flats avoid most of the problems associated with the halogen lamp but take much longer to expose (especially in the blue), and contain weak emission lines from the lamp and weak telluric absorption lines from air within the dome.  The dome lamp is not hot enough to calibrate the NUV range.

\begin{figure}[tbp] 
   \centering
   \includegraphics[width=5.5in]{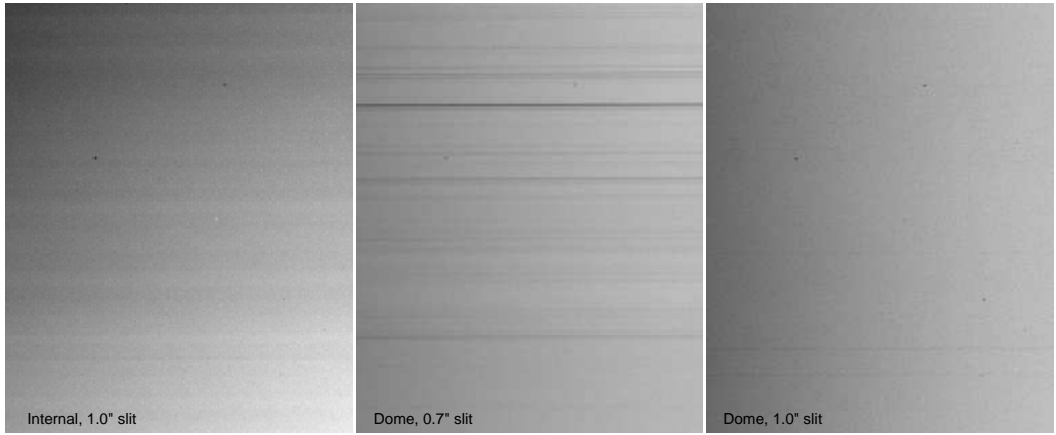}
   \caption{\small
   Comparison of dome and internal flat-field calibrations and of the 0.7$\arcsec$ and 1.0$\arcsec$ slits.  The halogen lamp shows significant spatial structure (banding) and an overall gradient that must be removed prior to flat-fielding.  The 0.7$\arcsec$ slit shows numerous defects (small obstructions in the slit).  Dome flats with the 1.0$\arcsec$ slit (or 1.5$\arcsec$ slit) provide the cleanest flat-fields. 
 }
\label{fig:flatfields} 
\end{figure}

\item Sky flats taken during twilight are in many ways superior to either dome or halogen flats; they are guaranteed to show (almost) no spatial nonuniformity and include more UV signal than is available from the dome or halogen lamps.  However, they can only be taken in a short window at the start or end of the night and are not obtained by most users.  The twilight spectrum shows many strong (Solar) absorption lines that must be modeled and removed to a high degree of precision for the observations to be useable as flat-fields.

\item Deuterium flats (an alternative form of internal flat using a deuterium lamp) are a relatively new addition to LRIS; a dense set of broad molecular lines produces a pseudo-continuum in the ultraviolet.  They are the only practical option to calibrate the ultraviolet on LRIS, but they contain substantial line structure and (like halogen flats) also exhibit non-uniform illumination along the slit direction.  They are suitable only for calibrating the blue camera.
\end{itemize}

When run in default mode, the pipeline will only construct the processed flat-fields that it needs for those science configurations that were actually observed that night (although it can be instructed to construct processed flats even for configurations with no science data as well).  If multiple flat types are available within a given configuration, the pipeline will produce separate processed flat fields for each flat category and choose between them later. 

Flat-field exposures are identified automatically by the pipeline based on their header keywords (e.g., dome flats can be recognized by checking the relative azimuthal offset between the telescope and the dome, which should be 90 degrees exactly), amount of continuum signal, and exposure times.  These are sorted into groups of exposures taken with the same configuration, which will be combined together to create a processed flat.  Exposures with low S/N are excluded automatically.

For flat types without strong line spectra (halogen/dome), the wavelength structure of each individual flat exposure is removed via a simple (if somewhat crude) means: each pixel (wavelength) column is divided by its median value.  For flats with strong line structure (sky/deuterium) this basic method will not work because of the slight ``tilt'' of the lines relative to the vertical axis of the CCD.  For these flats, a response curve at the middle of each CCD is calculated, then shifted to match each individual pixel row's wavelength offset and divided. 

No flat-field lamp presents sufficient UV flux to calibrate the far UV end of the blue camera, and most lamps are unable to calibrate the near-UV either: typical wavelength limits for producing usable counts are 4100 \AA\ for dome flats,  $\sim$3700 \AA\ for halogen flats, $\sim$3200 \AA\ for twilight-sky flats and 3000 \AA\ for deuterium flats.  The pipeline therefore does not attempt to flat-field data outside wavelength regions for which there is useable signal; it simply takes a wavelength average over the high-S/N regions and applies this to every row across the (low-S/N) UV.  Future pipeline releases may employ a standard pixel flat synthesized from a variety of nights and flat-types to avoid this issue and improve flat-fielding generally.

For halogen flats and deuterium flats, the spatial gradient and ``banding'' structure along the slit direction is removed by taking horizontal medians and dividing.  (This procedure is not necessary for dome or sky flats.)  The removal of this banding is accurate only to the level of about 1\%; the small residuals will have only a minor direct impact on the source itself but may lead to additional residuals associated with sky lines or twilight spectra in the final spectra for nights in which internal flats were used.

After the horizontal (and for halogen/deuterium flats, vertical) response corrections, the flat exposure sequences are median-combined together and saved to disk.  

\subsection{Flat-fielding}
\label{sec:flats}

In comparison to the construction of the flat-field, the calibration of science data using this field is straightforward, at least in principle: each science exposure is matched to an appropriate flat taken in the same configuration and divided (arc exposures are also flat-fielded although this is not strictly necessary).  However, there are some subtleties.

First, it is worth noting the issue of binning and cropping.  While most observers will use the same binning settings for the entire night, it is conceivable that an observer will procure some binned and some unbinned frames either deliberately or by accident, and not acquire flats in the same mode.  It is possible to rebin and crop frames in software to deal with this fact and LPipe will do this automatically when necessary.  Likewise, while many observers will choose to set the red CCD window settings to read out only those areas on the chip that are actually exposed to astronomical signal (to save disk space and transfer time), some frames may accidentally be acquired in a different window setting.  The pipeline keeps track of binning and cropping coordinates (both at the telescope and as applied by the pipeline during processing) and will ensure the frames are aligned appropriately, allowing e.g. unwindowed flats to be applied correctly to windowed data.   If science observations were taken in full-frame mode, it will also crop the image down vertically to only include the area that is actually exposed by the longslit to save read/write time.

Another important subtlety concerns irregularities in the slit, in particular for the 0.7\arcsec\ slit.  This slit has a large number of obstructions (``defects'') along the slit that block a small amount of light at certain positions, an effect that is clearly visible as fine horizontal banding in flat-fields taken with this slit (Figure \ref{fig:flatfields}).  The 1.0\arcsec\ slit has a much smaller number of irregularities (the 1.5\arcsec\ slit has none).  In principle, these variations could be removed from science frames by flat-field division.  This is not possible in practice without special calibrations and/or techniques: the physical location of the slitmask with respect to the CCD moves during the night in response to flexure, and thus the locations of these bands are observed to move vertically up and down by up to several pixels between different exposures.  While the slit variations are only a few percent, they are highly undersampled and this will have severe consequences on the quality of 2D sky-subtraction.  In fact, dividing a science observation by a flat with a slit-defect trace at the ``wrong'' location will produce positive+negative trace pairs which can then falsely be detected as objects at later processing stages.  Tools have been developed to mitigate these effects, but they are not automated within the current version of the pipeline and users should be aware that for the 0.7\arcsec\ slit in particular, unsupervised reductions are likely to produce science results that are significantly worse than what is possible in principle with properly flat-fielded observations.

Regions of the flat-field with transmission below 1\% of the peak (typically, bad pixels and portions of the CCD away from the longslit) are set to NaN in the final image to avoid immensely magnifying noise.

\subsection{Splitting}

After flat-fielding, all images are still stored as simple (single-extension) FITS files with the two chips merged together.  However, it soon becomes justified to split the chips again, for two reasons.  First, this saves processing time during the computing-intensive cosmic ray rejection and sky-subtraction steps for the (extremely common) case that only one chip is of interest.  Second, this avoids having to continue to recognize the spatial discontinuity at the slit gap position in future processing steps.  Therefore at this stage the two chips are split into separate files.

By default only the right chip is split off and the left chip is ignored from this point forward.  The user can override this default and retain both chips by specifying this option at the command line when the pipeline is run.  However, automated source identification and extraction is currently only available for sources on the right chip since in almost all cases the target of interest is centered on the right chip.

\subsection{Cosmic ray cleaning}

\begin{figure}[tbp] 
   \centering
   \includegraphics[width=7in]{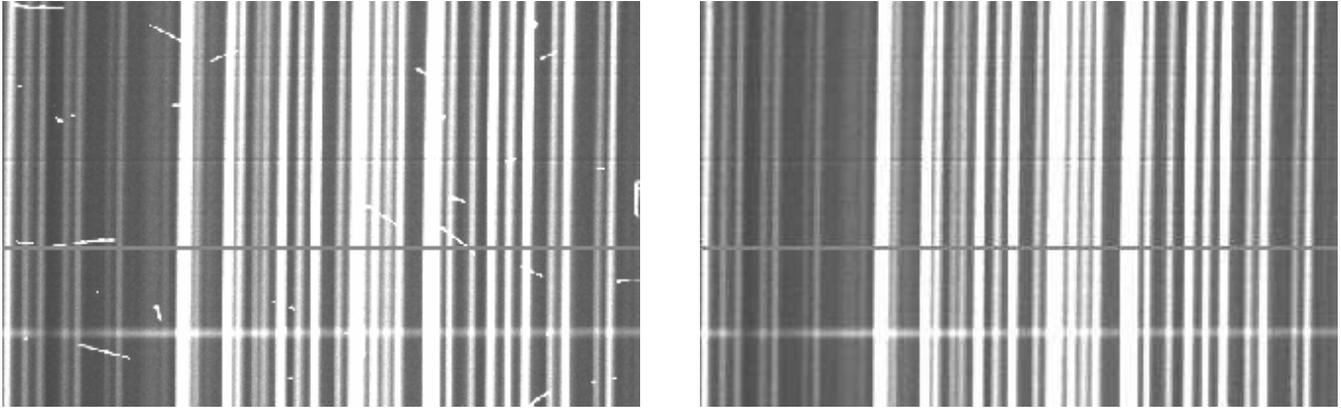}
   \caption{\small
   Cosmic ray cleaning demonstration.  While the upgraded LRIS-red detector is significantly affected by cosmic rays (left half), these can be efficiently detected and removed by a specialized algorithm (right half, shown after cleaning).   The wide horizontal grey band just below the middle of the images is a bad column; the faint horizontal depression in the flux just above the middle is due to a slit defect (\S \ref{sec:flats}).
 }
\label{fig:cosmicrays} 
\end{figure}

A casual glance at a long exposure on LRIS-red immediately reveals the challenges involved in processing data from this detector: the chip is filled with long charge trails from radiation and cosmic-ray hits, affecting a significant fraction (up to 1\%) of the pixels per exposure (Figure \ref{fig:cosmicrays}).  The trace of the target source is likely to be affected at several points.  This is a common feature of deep-depletion CCD arrays.  

Old cosmic ray detection algorithms (such as the basic tool in IRAF) work poorly on such features because they affect so many pixels over a large area.  Tools developed for space-based imaging may also struggle due to the presence of the bright vertically-structured sky lines which may themselves be moderately undersampled in some binning modes.  Median-combination methods are unsuitable for variable sky conditions and require at least three exposures, which are not obtained by most observers due to the associated read-noise and readout-time penalties.  

For these reasons a custom tool was developed for removal of cosmic rays for deep-depletion 10-m-class spectra.  This algorithm proceeds in several steps: first, it employs a row/column median subtraction to remove sky lines and source continuum; next, it attempts to model the value of each pixel in the image based on groupings of its neighbors and identify cosmic rays as pixels whose model values disagree both with each other and the real value; finally it employs a percolation algorithm to try to identify every pixel affected by a long particle trail.

The procedure is not perfect.  Cosmic rays with directions of travel along the spatial direction are usually missed because they cannot be distinguished from sky lines using the current algorithm.  The percolation procedure sometimes misses a few affected pixels.  Most alarmingly, binned observations of sources with strong, narrow emission or absorption lines can cause these lines to be misidentified as cosmic rays in some instances.  In general, observers doing narrow-line science should employ caution in verifying the pipeline products after this step.  Observers with no interest in emission lines (and not employing more than 2$\times$2 binning on the red or 2$\times$1 binning on the blue side) could safely make the pipeline more zealous and improve the cleaning efficiency. 

After cosmic rays are identified they are filled in with a median-model estimate of the pixel values.  A mask image indicating the positions of affected pixels is retained for use in flagging compromised regions of the spectrum in later reduction stages.

The blue side is much less-affected by cosmic rays but still contains them; a simplified (and faster) version of the cosmic ray cleaner is applied to blue observations.

Lamp spectra are not cosmic ray cleaned:  these are always very short exposures\footnote{Except when using the Fe lamp, which is ignored by the pipeline} and cosmic rays can be removed by using median-averaging during the extraction.

\subsection{2D sky subtraction}

The next step is to 2D sky subtract the images in order to separate signal originating from Earth's atmosphere (of interest only for calibration purposes and to estimate the noise level that results from them) from the signal originating from astronomical targets (the actual sources of interest).  

The basic underlying procedure is to take medians along each wavelength column and subtract them.  However, a simple median would not be able to deal with even a small amount of curvature or tilt in the sky lines, so in actuality the medians are calculated in blocks, interpolated, and then fit with a simple local regression model.  The sky flux is taken out of the science image and written to an extension.

This algorithm works well for modest tilts but fails if the sky lines are tilted with respect to the CCD grid by more than a few degrees, which can be problematic for some gratings.  Furthermore, although smooth continuum from sources along the slit are removed by a separate median procedure before fitting the sky lines, absorption and emission lines will subtly influence the sky model and typically result in under/oversubtraction of the sky around them, respectively.  Also, very large sources (nearby galaxies) will have any low spatial-frequency components removed as ``sky''.  Fortunately, this is not generally a problem:  the only function of 2D sky subtraction when running the pipeline in default mode is to remove enough sky signal to identify and trace faint sources in later steps, and over- or under-subtraction is unlikely to affect this significantly.

An alternative sky-line subtraction routine based partially on the method of \cite{Kelson+2003} has also been developed and integrated into the pipeline.  The performance of this implementation is not yet as reliable as spatial median subtraction under most circumstances\footnote{Helpfully, most LRIS gratings/grisms have minimal `tilt' relative to the CCD $y$-axis, so traditional methods work relatively well.  Median-averaging is also more robust to complicating factors such as flat-fielding problems (\S \ref{sec:flats}) or bright/variable galaxy backgrounds.}, but the two algorithms are interchangeable within the codebase and the user can choose which to employ at runtime.

A different type of alternative method for sky-line subtraction employs nodding (dithering the telescope back and forth along the slit, and subtracting).  Nodding is not commonly used by LRIS observers because of the long readout times and the competing demands of the blue spectrograph (which is read noise limited for faint-object observations), although it was much more commonly employed in the past in order to mitigate the severe line fringing in the near-infrared observed with the old red detector.  Support for nodded pair subtraction does not currently exist in the pipeline but may be added at a later date.

\subsection{Shifting and stacking}

It is common, especially on the red side, to acquire repeated exposures for a single object.  It is beneficial to stack these prior to source extraction, since this makes it much easier to identify and trace faint sources.  Therefore the pipeline looks for a sequence of consecutive exposures with the same configuration and at the same position on the sky \emph{except} for small dithers along the slit direction.  Using the header positional keywords, all frames after the first are shifted vertically to match the spatial location of the initial exposure (if necessary), and then simply summed.

In some cases the user may want to extract multiple observations of a source as individual spectra - for example, in order to obtain a time series for a source that is rapidly varying, or as a result of variable seeing or transparency.  If a large number of short exposures are taken in sequence of the same object the observation is assumed to be a time series and in such cases the exposures are not summed; the user can also request certain targets (or all targets) not to be summed with an appropriate command.

In other cases, a temporary guiding loss or other error may cause repeat exposures in the same setup to not be summed, or the user may want to combine observations of the same source taken at different times in the night.    These can be summed explicitly by the user in 2D space or later in 1D space using the appropriate individual commands.

\subsection{Tracing}

Tracing (determining how the wavelength-dispersed continuum signal from an object tilts and curves as it extends along the CCD in the $x$ direction) is conducted prior to (and separately from) extraction.  This aids the identification of very faint sources, and avoids having to repeat the tracing procedure if the aperture extraction details are changed later.

For bright sources (such as standards) tracing is straightforward: the spatial ($y$) centroid is calculated for every wavelength ($x$) column and a fitting function ($y = F(x)$) is applied to calculate a smooth trace curve.  This is not always reliable for science observations where the source may be weak at some or all wavelength columns.  To improve S/N, the observations are (temporarily) binned up by 32 pixels or more along the wavelength axis, producing a series of spatial profiles along the (entire) slit at different wavelengths.  These profiles are then cross-correlated to measure the spatial offset---starting in the middle of the wavelength range where sensitivity and source counts are likely to be maximized, and then progressing outward in both directions.  A polynomial fit is applied to the resulting offset array to determine the trace function, the parameters of which are stored on disk.  If a consistent, slow-varying trace function cannot be found from this method because there are no sources apparent in the image then no trace file is written (this is not a problem: subsequent steps will simply rely on trace files from other, brighter targets.)

\subsection{Source and aperture identification}

Next the pipeline must identify what sources to extract.  This is the most likely step where a user may want to intervene in the automated pipeline operations (generally, to modify apertures or background subtraction behavior, or to extract a different source).

First, the pipeline completely collapses each image along the wavelength axis (the true wavelength axis as determined by the trace solution determined above, not the CCD $x$-axis) to produce a measurement of the 1D spatial profile along the slit.  Significant peaks in this profile (candidate sources) are then identified.  The pipeline chooses which of these is most `likely' to be the primary science target based on proximity to the pointing origin (as measured by the location of standard-star traces), with the brightness of the source as a tiebreaking factor if multiple sources close to the pointing origin are detected.

This routine is initially carried out independently for blue and red data.  In principle this could lead to a different object being automatically extracted on the red vs. the blue side (and these spectra would then be fused together into a chimeric spectrum in future steps). To avoid this, the pipeline cross-correlates the blue spatial profile and the red spatial profile for each pair of time-overlapping observations to determine the astrometric offset, and checks that the object centroids are consistent within a few pixels.  If not, one of the two apertures is changed to ensure consistency with the other.  If the difference is small the two positions are merely averaged together.  Aperture diameters are also averaged for consistency.  (The red-blue offsets of standard-stars observed in the same configuration are used as a safety check to ensure that the cross-correlation procedure does not falsely identify the astrometric offset.)

Because the entire wavelength axis is summed over, this procedure can effectively identify sources as faint as 25th magnitude (in long exposures under dark-sky conditions).  Sources with continuum fainter than this, including objects dominated by emission lines, may not be recognized; these will have to be pointed to manually by the user (see below).  

In any case, no automated routine can determine with certainty which source(s) a user is interested in or the optimum way to extract them.  While extraction of bright stars and QSOs is usually obvious and intervention is rarely needed, for more subtle cases (a faint star in a crowded field, a supernova embedded in a bright galaxy) it is often necessary to specify exactly which source to remove and how to deal with the surrounding background.  

\begin{figure}[tbp] 
   \centering
   \includegraphics[width=6.8in]{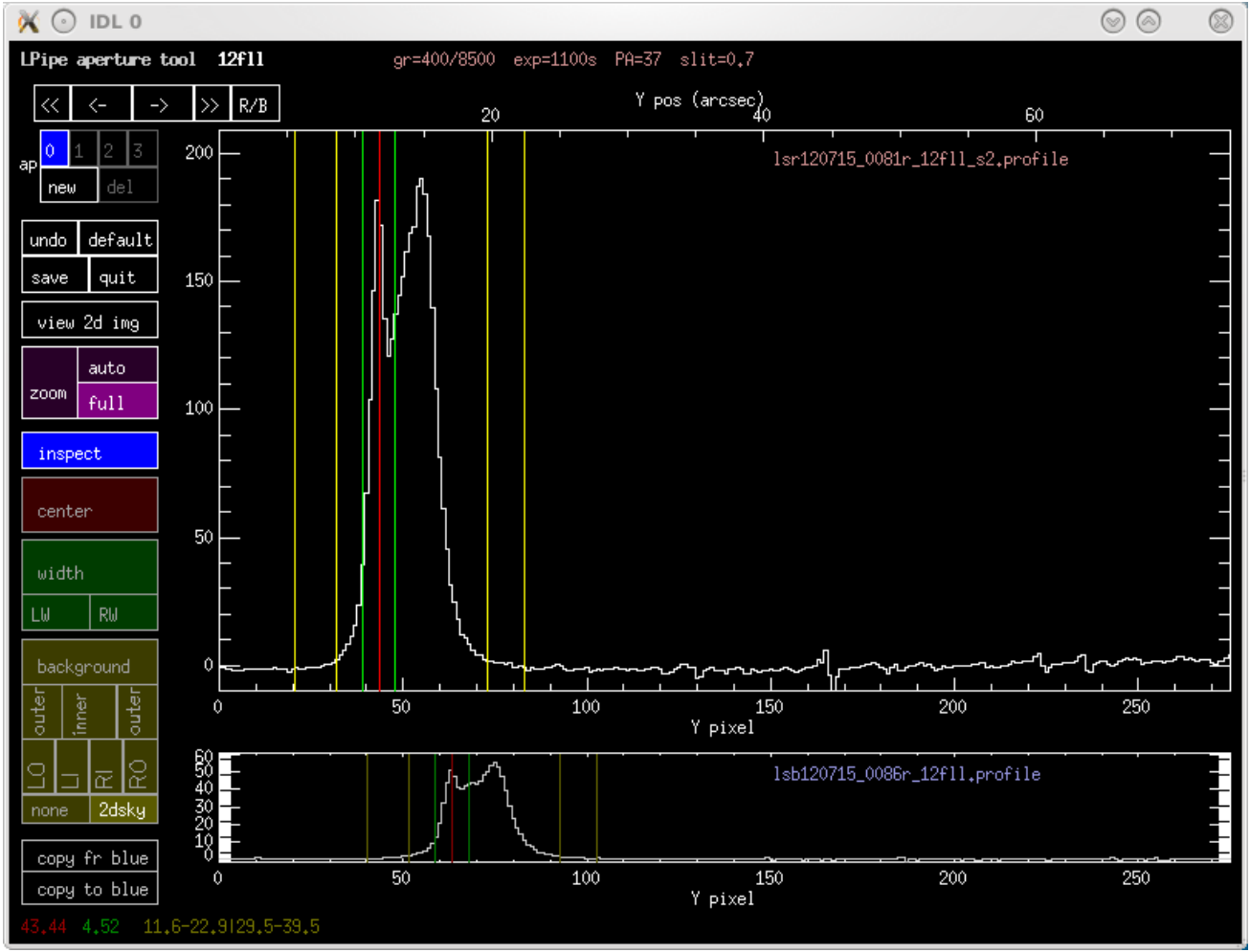}
   \caption{\small
   Screenshot of the aperture selection tool GUI available in the pipeline, showing an object's spatial profile along the slit, the selected aperture region (green), and background-subtraction regions (yellow).  Apertures can be resized, moved, added, or subtracted, and the background subtraction algorithm can be changed interactively by the user.  This is necessary only for objects with close neighbors or in complex regions; in most cases the automated trace identification can be relied on. 
 }
\label{fig:setaperture} 
\end{figure}

As a result, a graphical interface for aiding source selection has been developed to supplement the pipeline (Figure \ref{fig:setaperture}).  This tool enables the user to select the source (or multiple sources) to extract visually and contains options for changing the aperture width, background locations, and other aspects of the extraction.

\subsection{Extraction}

The actual extraction of a source is straightforward since all details of the aperture placement have been determined in advance.  The pipeline does not rely on the 2D sky subtraction for this step due to its risk of accidentally removing source flux with sky flux: the sky and source extensions are combined together again in memory before all extraction operations.

Simple extraction is used to obtain a measurement of the sky-unsubtracted source flux for each pixel column:  each column of the 2D spectrum is convolved with a boxcar extraction kernel and summed.

To remove the sky background, each column is individually median-filtered (to help remove any cosmic rays or artifacts missed by the cleaner) and the average flux is measured for both an upper and lower background region near the extracted source.  The two values are then interpolated to obtain the average sky flux over the object extraction aperture.

This procedure differs from the so-called ``optimal'' extraction method summarized in \cite{Horne1986}.  Optimal extraction requires very precise measurement of both the spatial profile and the trace function.  Both of these are supplied by the pipeline, but for faint sources or sources with very red or blue colors they may not be reliable.  Use of optimal extraction (especially for bright sources) will be explored in the future.  This could provide a S/N improvement of 40\% or more.

The output is saved to a simple ASCII file with several columns: wavelength pixel, source counts, sky counts, trace centroid, and the fraction of extracted pixels which were bad.  A FITS-like header containing all the key information from the original FITS header along with some information about the extraction parameters used is written at the top of the file.  This format is used for all 1D spectra written by the pipeline.

\subsection{Lamp wavelength calibration}

\begin{figure}[tbp] 
   \centering
  \begin{minipage}[b]{0.47\textwidth}
    \includegraphics[width=\textwidth]{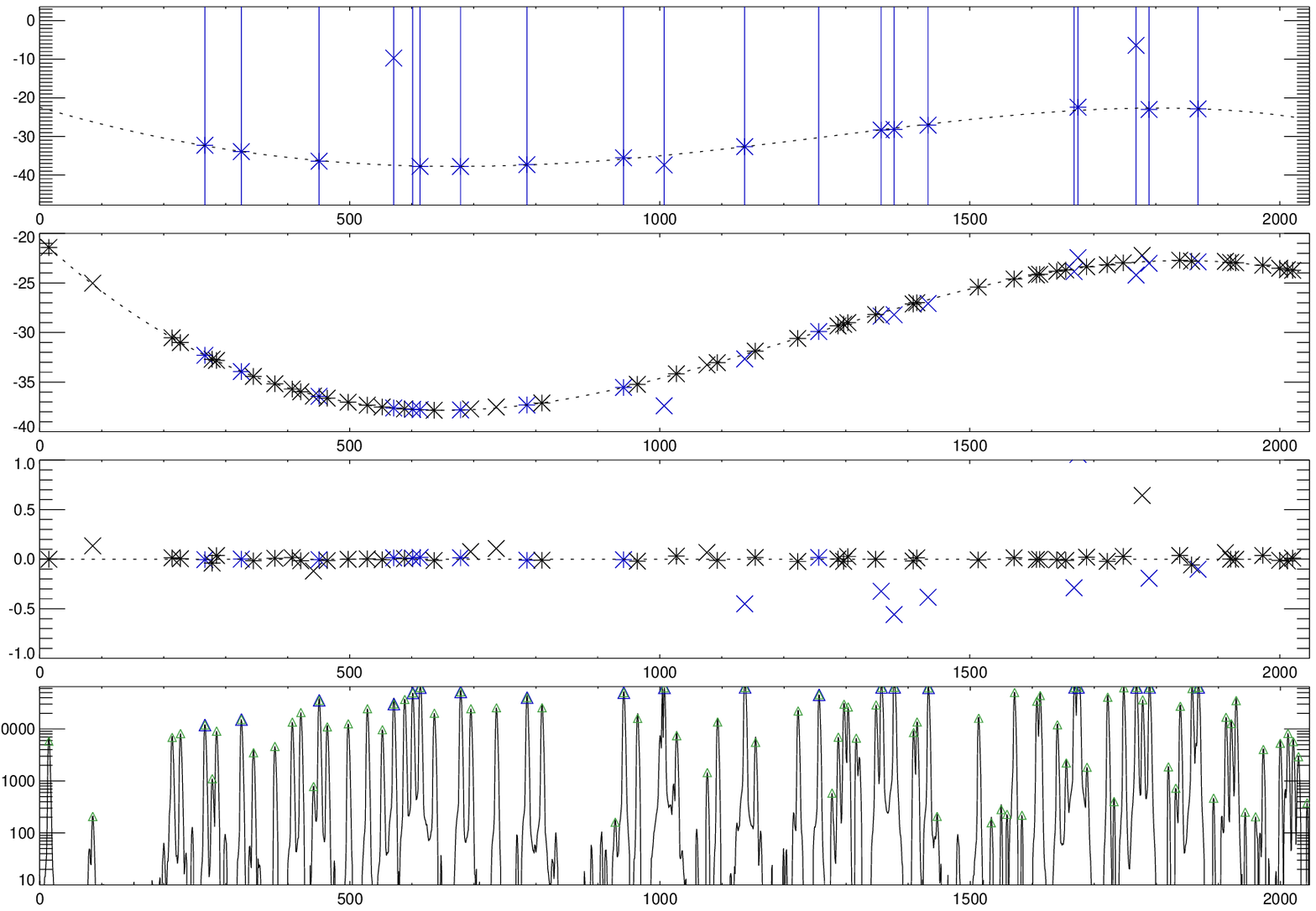}
  \end{minipage} 
  \begin{minipage}[b]{0.50\textwidth}
    \includegraphics[width=\textwidth]{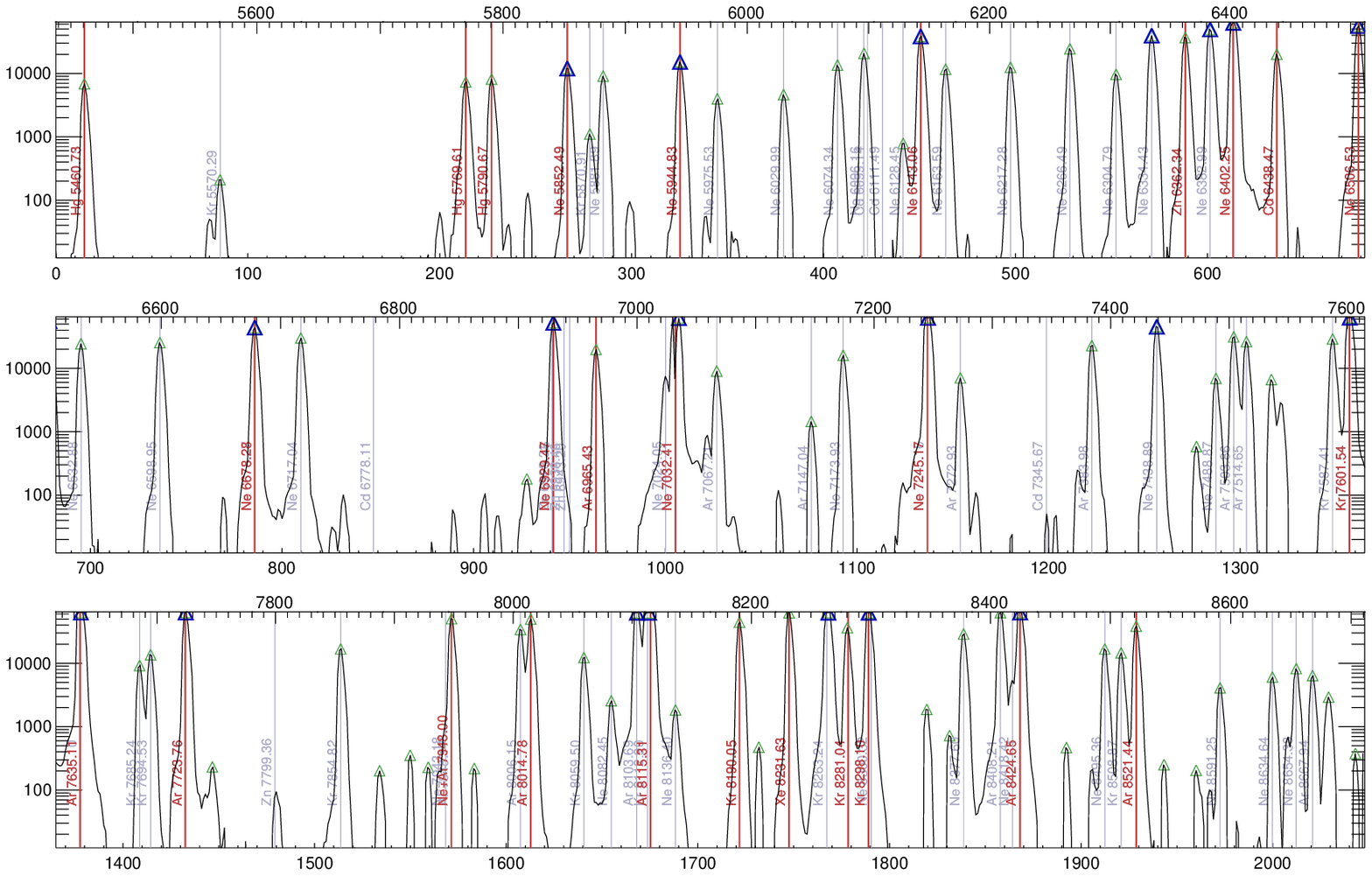}
  \end{minipage}
   \caption{\small
   Wavelength check plots for a typical low-resolution arc solution.  \textit{Top left:} Deviation (in \AA) from a purely linear wavelength solution for bright lines matched to bright reference lines, used in the initial low-order solution guess.  The x-axis is the column number. \textit{Upper middle left:} Deviation from a purely linear wavelength solution for the final solution, with all matched lines marked.  X's indicate line matches rejected from the final solution.  \textit{Lower middle left:} Deviation from the final wavelength solution for all matches.  \textit{Bottom left:} Arc spectrum (in counts; log scale).  \textit{Right three panels:} Detailed arc spectrum, with bright reference line wavelengths marked as red vertical lines and faint reference line wavelengths marked as gray vertical lines.
 }
\label{fig:arcs} 
\end{figure}

For maximum flexibility (to allow new gratings, CCD changes, etc) the pipeline treats wavelength calibration in a fully general manner: no archival wavelength solution is used and the wavelength-to-pixel solution is treated as a novel problem with no strong priors other than the approximate initial central wavelength and dispersion, plus the degree of polynomial to fit.

Basic extraction of all valid comparison lamp spectra (loosely referred to as ``arcs'') is first performed (a vertical median is taken over a limited region close to the standard trace position) to produce 1D spectra, which are then extracted using a peak detection and centroiding algorithm to produce a line list of pixel positions.  A custom pattern-matching algorithm is used to identify matching groups of bright line triplets in the observed line list and the reference line list.  (Candidate match triplets are evaluated based on the ratio of wavelengths within the triplet: i.e., three observed lines at pixels $x_a$, $x_b$, $x_c$ are considered as possible matches to reference wavelengths $\lambda_1$, $\lambda_2$, $\lambda_3$ if $\frac{x_c-x_b}{x_b-x_a} \sim \frac{\lambda_3-\lambda_2}{\lambda_2-\lambda_1}$.  An individual observed line must be matched consistently to a reference line in many triplet-pairs to be treated as a genuine match.)  At that point a linear regression model is used to fit the wavelength solution, which is successively refined by dropping lines that are poor matches (these may be saturated lines, blended lines, or misidentifications from the pattern-match) and re-matching until a good solution is achieved.  A high-order polynomial is needed to accurately fit the whole wavelength range for low-resolution setups (currently a 5th order polynomial is used for the blue and a 7th order polynomial for the red).

A practical challenge is the fact that the LRIS arc lamps have very few lines around 5500 \AA (the only bright line between 5085--5769 \AA is the Hg line at 5461\AA).  This region includes the wavelength limit of the blue 400/3400 and 600/4000 grisms, and will likewise include the limit of the red grating if continuous observations are desired.   If the D560 dichroic is used, this is also a region of rapid variation in the transmission on both cameras and small wavelength inaccuracies can impart large flux calibration inaccuracies, which in turn will lead to inaccurate red-blue relative scaling in later phases (ultimately producing a jump in the flux at the dichroic wavelength).  Users should be alert for flux- and wavelength-calibration artifacts in this region.

Good, precise wavelength solutions require as many lines as possible: when using the D560 dichroic, blue calibration requires the Hg, Cd, and Zn lamps at minimum and red calibration requires the Hg, Ne, and Ar at minimum.  (Both sides benefit from even more lamps being on, and it is generally optimal for the entire suite of arc lamps to be illuminated for all calibrations.)
However, the lengthy warm-up time for the Cd and Zn lamps often leads to exposures being taken too early before these lines have appeared.  These ``unwarmed'' arcs will be inadequate to solve the wavelength solution, especially for the blue side.  If multiple identical arcs are taken the pipeline will use only the last one (under the assumption that earlier lamps are more likely to be unwarmed).  However, wavelength solutions are always validated internally to check that all expected bright lines are present, so even if ``bad'' arcs are present they will be ignored in further processing.

Successful solutions are written to disk in the form of a column of coefficients from the polynomial fit.  Every available arc exposure is solved individually, except when several identical arcs are taken in series (see above).  Decisions about which of these solutions to employ to calibrate which science exposures are taken later.  The pipeline is able to translate solutions derived from arcs taken in one binning to science frames taken in a different binning, so it is not necessary for the user to acquire arc calibrations in the same binning as science observations.\footnote{While most users will acquire all observations in a consistent binning setting, acquiring arcs in 1x1 binning mode (regardless of the science setting) can actually be useful to avoid saturation of the brightest lines for binned observations with low-resolution gratings.}

Residuals of the wavelength solution are typically $\sim$0.02 pixels for the red camera and $\sim$0.15 pixels for the blue camera.  However, flexure correction and the different illumination pattern of the sky versus the reference spectra typically limit the real-world wavelength-calibration accuracy to 0.2 pixels for both sides (and occasionally worse if flexure correction cannot be performed: see next section).

\subsection{Science wavelength calibration and flexure correction}
\label{sec:wavcal}

Next, the pipeline will apply the wavelength solutions derived from the arcs to the extracted science frames.  This requires matching each science frame to an appropriate (solved) arc frame and applying a subsequent flexure correction.

For each configuration, LPipe will look up an appropriate global reference arc exposure (usually one from late in the afternoon) and read in its wavelength solution.  Only this arc will be used to estimate the higher-order (dispersion and curvature) terms of the wavelength solution in order to guarantee that all observations taken using the same setting use the same original arc calibration file.

Due to flexure, adjustments to this wavelength solution will be necessary to calibrate observations taken during the night.  In most cases the night-sky emission lines in the spectrum can be used: a line list is estimated from the night-sky spectrum and matched to a catalog to determine a linear offset.

If this fails or is impossible due to a short exposure or bright twilight, other methods are employed.  If an arc was taken at the precise position of the observation, the zeroth-order term of that arc is substituted for the one from the reference arc.  If the object was very bright (e.g. a standard star), the region of the spectrum around the telluric A band (or the B band if A is not covered) is cross-correlated with a reference telluric spectrum to estimate the wavelength offset.

There are no bright sky lines at wavelengths shorter than 5577\,\AA\ (one weak line is present at 5197\,\AA\ but its S/N is always low), and no telluric lines either.  As a result, flexure correction is often difficult for LRIS-B.  The 5577\,\AA\ OI line is always present in the wavelength coverage of the 300/5000 and 400/3400-line grisms, and this line alone is usually enough to measure the flexure correction for observations taken with these grisms.  The same line is normally present in the 600/4000 grism as well, \emph{but} a hardware intervention in 2012 resulted in a shift in wavelength coverage to the blue that meant that the line shifted off the CCD until the D560 dichroic angle was changed in early 2014.  Observations with the 1200/3400 grism will never show this line, nor will observations with any grism if the D460 or D500 dichroic is used.  In these circumstances on-sky arcs currently provide the \emph{only} way of providing an accurate wavelength solution.\footnote{It may be possible to use cross-correlation of the sky continuum against a night-sky spectrum (in dark time) or Solar spectrum (in bright time)} to correct blue observations in the future, but this feature does not currently exist within the pipeline.

\subsection{Flux calibration}
\label{sec:fluxcal}

Flux calibration involves calculation of the response function (sometimes called the sensitivity function) using an object whose intrinsic spectrum is known.  In its simplest form, the response function is described by:

$$C(\lambda) = F_\lambda(\lambda) R(\lambda)$$

Here, $C$ is the count rate in ADU/s (per extracted pixel) as calculated earlier\footnote{The ADU values are corrected for flat-field pixel variations but \emph{not} gain or illumination variations.}, $F_\lambda$ is the true intrinsic spectrum of the source, convolved to the resolution of the instrument, in erg s$^{-1}$ m$^{-2}$ \AA$^{-1}$, and $R$ is the response function (units: ADU erg$^{-1}$ cm$^2$ \AA), which allows us to convert back and forth between these two quantities.  

\begin{figure}[tbp] 
   \centering
   \includegraphics[width=5.5in]{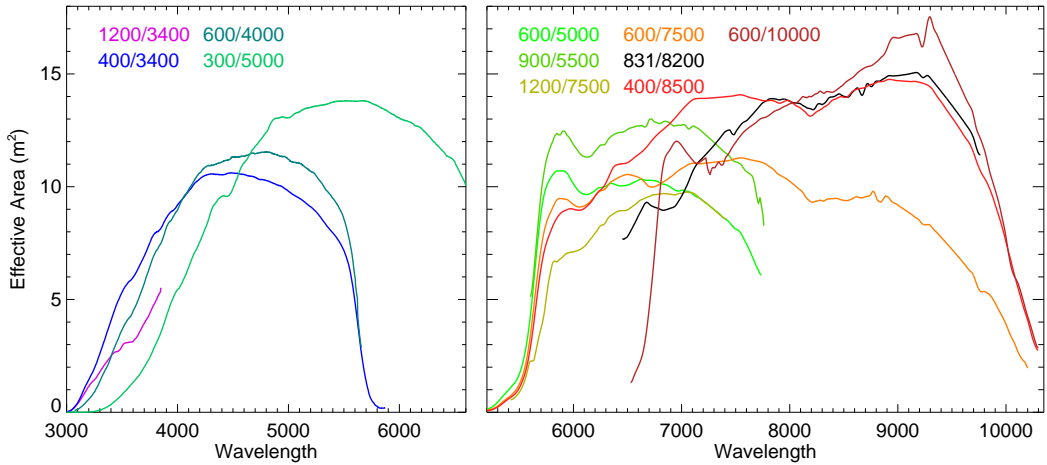}
   \caption{\small
   Effective area and total system efficiency (including atmospheric and slit losses, except telluric absorption) of LRIS-B (left) and LRIS-R (right) for a variety of dispersers, based on a median combination across one year of standard-star observations between September 2016 and August 2017.  All curves are for the D560 dichroic and no filter, except for the 1200/3400 and 300/5000 grisms (which used the D500 and D680 dichroics, respectively) and the 600/10000 grating (which used the D680 dichroic).  Narrow features are generally artifacts introduced by intrinsic lines or the median combination procedure.  Because losses include atmospheric effects and different curves are based on different nights, the absolute scalings between curves are not necessarily indicative of a true difference in efficiency.
 }
\label{fig:effarea} 
\end{figure}

It is simplest to assume that $R$ is a function of wavelength alone and thus that it can be calculated in a simple manner as $R = C / F$.  In reality, however, it is a function of detector position and wavelength.  Most pixel-to-pixel variations were removed by flat-fielding, but broader variations associated with the detector or the illumination pattern cannot easily be solved for without a model of the spectrum of the flat-field source.  The distinction is unimportant \emph{as long as} the wavelength solution is the same for all spectra, such that pixel and wavelength can be treated as interchangeable.  This is approximately true in practice if calibrations are taken in the same setup as science observations, but is not exactly true in reality due to the effects of flexure.  We assume that the illumination of the CCD is constant within the flexure variation scale.  

As a first step, we solve for $R$ for every standard individually by simply dividing the reference spectrum of an object (interpolated to the wavelengths of the CCD pixel grid) by its observed count spectrum.  Reference spectra are taken from the CALSPEC database at STScI\footnote{\url{http://www.stsci.edu/hst/observatory/crds/calspec.html}}, the ESO libraries\footnote{\url{https://www.eso.org/sci/observing/tools/standards/spectra.html}}, and the ING compilation of standards\footnote{\url{http://www.ing.iac.es/Astronomy/observing/manuals/html\_manuals/tech\_notes/tn065-100/workflux.html}} and come from a variety of individual sources \citep{Oke+1983,Oke1990,Stone1977,Hamuy+1992,Hamuy+1994,Bohlin+1995,Bohlin1996}.  This is only an initial estimate of $R$: it contains noise (especially for short exposures and in the UV), and the limited resolution causes the observed spectrum to be `blurred' slightly on the detector (additionally, many library reference spectra omit the stellar absorption lines entirely!).  

These effects must be corrected for.  LPipe first looks up what wavelength regions of the spectrum are affected by lines, omits these from both the observed and the intrinsic spectrum, and interpolates over them.  Telluric absorption lines are initially interpolated over for the same reason (these will be dealt with with a different procedure momentarily).  The response curve is smoothed by Savinsky-Golay convolution to reduce noise.  The telluric absorption is then estimated by comparing the actual counts over the A, B, and water vapor bands to the interpolated counts over these regions: i.e., the pipeline calculates $a_{\rm tel}(\lambda) = C_{\rm obs}(\lambda) / C_{\rm interp}(\lambda)$.  This is required to be between 0 and 1 (any pixel values $<0$ and $>1$ are set to 0 and 1, respectively).  The non-telluric and telluric components (along with the wavelength and pixel values) are written to disk in an ASCII file.

A helpful byproduct of the response calculation is an effective area curve for the observation, which describes the equivalent collecting area for a perfectly efficient instrument located above Earth's atmosphere that faithfully records every photon incident upon its surface; this provides a direct measure of the real-world performance of the observatory.  It can be calculated as 
$\mathcal{A} = R G h \nu / \Delta\lambda$, where $G$ is the CCD (inverse) gain, $h \nu$ is the photon energy, and $\Delta \lambda$ is the wavelength range for one pixel; typically this is observed to be $\sim$10 m$^2$ in the blue and $\sim$15 m$^2$ in the red for observations in clear weather and average seeing (over the regions of optimal sensitivity of each grism and grating).  The ratio of this to the physical area of the primary mirror (73.3 m$^2$) gives the total throughput of the system as a function of wavelength, approximately 12--20\%.  This is consistent with the expected losses: a breakdown (at 6000\,\AA\ under typical conditions, given values from \citealt{Oke+1995}) is:  0.9 (atmosphere) $\times$ 0.86 (primary) $\times$ 0.88 (secondary) $\times$ 0.8 (slit/aperture loss) $\times$ 0.95 (dichroic) $\times$ 0.4 (rest of spectrograph) $\times$ 0.8 (detector QE).

Telluric absorption, atmospheric extinction, and slit losses all vary as a function of sky location and atmospheric conditions, and thus the response function observed for a standard may not be the same as the response function for a science object.  Slit losses, which may be wavelength dependent (seeing is generally worse in the blue), are difficult to correct for and provide by far the largest uncertainty in both relative and absolute calibration of LRIS data\footnote{This problem is greatly exacerbated if a the slit is off-target by an amount comparable to the slit size or if the focus is poor.}; the pipeline is not able to address this at present.  The other two effects (atmospheric extinction and telluric absorption) can be addressed more easily.   

The absorption properties of the atmosphere change gradually from night to night (due to variations in water vapor density or the amount of dust and aerosols in the atmosphere) and in principle it would be desirable to solve for atmospheric extinction using the nightly data, e.g. by observing multiple standards at different airmass.  Unfortunately, slit loss variations (which also vary with seeing, focus, and pointing) make this not reliable in practice.  Instead, differential atmospheric extinction is treated by employing the mean Mauna Kea atmospheric data from \cite{Buton+2013}, which describes the average opacity at zenith.  A polynomial is fit to their data to produce a generic zenith attenuation function, expressed in terms of optical depth as $\tau_{\rm atm}(\lambda)$.  This is used to convert the measured response function at the airmass of the flux standard to a predicted response function at the airmass of the science target: i.e. $R_{\rm cont, sci} = R_{\rm cont, std} \times {\rm exp}(-\Delta A * \tau(\lambda))$, where $\Delta A$ is the difference in airmass $A$ between science and standard targets ($\Delta A = A_{\rm std}-A_{\rm sci}$).  Fortunately the opacity of the Mauna Kea atmosphere is quite stable.  While variations of 10\% are observed in the relative opacity of the blue and the red, these are much smaller than errors associated with slit loss effects.

Telluric correction follows a similar procedure and is carried out effectively simultaneously with the overall flux calibration using the same standard star or stars (separate telluric standards are not yet recognized).  The primary difference is that there is no smoothing applied to the measured telluric response curve, and airmass rescaling is determined empirically: given an airmass difference $\Delta A$ between the standard and science target, the pipeline rescales the telluric attenuation factor according to the equation $f_{\rm tel,sci} = f_{\rm tel,std}$\^{}$(1 + \Delta A /A_{\rm std})$.   

A good estimate of telluric absorption requires a good estimate of the continuum outside the telluric bands.  If the standard is noisy or has weak intrinsic features that are not fully removed this may not be the case.  Furthermore, this telluric procedure does not take into account the limited resolution of the instrument.  In practice, this limited resolution makes the true airmass correction more complicated than the above equation describes outside the optically-thin limit due to saturation effects.  Additionally, because resolution is seeing-dependent, the appropriate telluric correction in practice will depend on seeing.  These two effects will inevitably lead to some residuals, although these are usually small in practice as long as standard-star observations are carried out at similar airmass ($\pm 0.3$) to each science target.  If science observations are taken at very different elevations from any standard star, or if seeing (or focus) vary greatly, this will inevitably cause large residuals over the telluric bands.

The pipeline must choose which standards to use to correct which science targets.  It will, of course, will only consider standards observed in the same instrumental setup as the science observation being calibrated.  If multiple standard-star observations are obtained (generally a good idea), it must choose between them.  The primary consideration in choosing between standards is the airmass difference versus the science target.  However, many other properties are also considered, including:  whether the standard star is known to have a smooth continuum (vs. many absorption lines that may not be well-characterized in the internal database), whether the standard exposure has a good wavelength/flexure calibration (either a long enough exposure time to derive sky lines, or an on-sky arc), whether the standard was observed close in time to the science target, and whether the flexure difference is observed to be small.  The pipeline weights these factors together to decide on the ``best'' standard to use.  If multiple good standards exist, two standard response functions may be averaged together.

An important complication that occasionally arises is that of cross-contamination of parts of the spectrum by light at other wavelengths, due to second-order light or reflected light.  In most LRIS setups, this is avoided by the use of the dichroic, which cuts off light from orders other than the one desired.  However, the dichroic does transmit some blue light and reflect some red light. These effects are currently neglected by the pipeline as no observable impact has been seen in any standard setup, but caution should be exercised when observing very blue or very red targets or in nonstandard setups.

One particularly important case that is commonly encountered in practice involves the use of the 300/5000 blue grism with the D680 dichroic.  In this setup the blue spectral region longward of $\sim$6000\,\AA\ is severely contaminated by second-order light and/or reflections.  
The pipeline will reduce data from this setup (truncating the wavelengths most severely affected by contamination) but the output should be interpreted only with caution since both the target and the flux-standard used to calibrate it will suffer from poorly-quantified (but typically large) contamination issues.

A practice employed by some observers to mitigate against light from unwanted orders/reflections (even in setups where it should be minimal) is to observe hot, blue stars to calibrate the blue side and to observe cool, red stars to calibrate the red side.  Unfortunately, cool stars usually also have more prominent atmospheric absorption features than the hot, blue white dwarf stars generally preferred as standards.  Currently the pipeline exhibits a preference for blue standards with few lines even for the red side and red-specific standards of this type are likely to be ignored in reductions if a hotter standard is also available, but the user can specify specific calibration targets if desired.  

\subsection{Red/blue connection}

Up to this point the red and blue cameras have been treated as if they were entirely separate observations conducted with different telescopes (with the exception of a concordance check to be sure the same source is being automatically extracted).  However, most observers want a single file describing the combined spectrum of each source.  Therefore the two spectra have to be combined.

As with many pipeline steps, this seems like it should be trivial but can be complex in practice.  The red and the blue sides may have been calibrated with different flux standards, with different slit/aperture losses and transmission losses, introducing an absolute (multiplicative) offset between the red and blue science spectra.  (Offsets are sometimes seen even when the same standard is used in the event that the tracing or trace alignment is imperfect.)  This must be removed to avoid imposing an unphysical discontinuity and to avoid turning errors in the absolute calibration (rarely a problem for most scientific purposes) into errors in the relative calibration (which are a problem for many applications).

This procedure is only possible for observations in which there is overlap between the red and blue spectra.  (If this is not present, the two files are simply merged together with a gap and the final spectrum is written.)  The pipeline uses the relative response curves (\S \ref{sec:fluxcal}) to identify an overlap wavelength region where there is acceptable sensitivity on both cameras.  

For bright sources, the pipeline then calculates the flux ratio between the red and blue within this region at each pixel, then determines a statistically-weighted average, which is the overall scaling factor.   By convention, whichever camera shows the lower transmission level is scaled upwards, leaving the flux scale of the spectrum from the other camera unaffected.   (Sky, uncertainty, and response values are also rescaled).  

For objects with very faint continuum this procedure is not possible because the S/N is too low.  In principle, the sky continuum level or (for the D560 dichroic) the total sky flux under the OI 5577\AA\ emission line could be used, although this requires accurate tracing on both sides right up to the CCD edge as well as very good bias subtraction, and in practice this technique is not sufficiently reliable for pipeline use.  For now, if the S/N is judged to be too low for a direct rescaling (or if there is no wavelength overlap) no scaling is performed: the spectra are joined together at the junction point with no scaling.  In such cases the user may wish to carefully examine the spectrum and perform a manual rescaling by eye as needed, or perform independent calibrations of the blue and red by comparing synthetic photometry of the spectrum in a standard filter passband to external photometry of the source.

In principle it would be possible to co-add blue and red spectra over the overlap region using a weighting function.  This would involve interpolation/resampling, which in general the pipeline avoids in order to keep the noise properties as simple as possible, and the S/N would be improved only over a narrow wavelength region where the transmittance in both directions (blue and red) is similar.  Instead, the pipeline calculates using the response curves a junction point where both spectra are similarly attenuated by the dichroic and joins the spectra there.  Note that when the 600/4000 blue grism is used or if the red grating angle is set to have minimal overlap, this point may be at the very limit of the blue CCD or the red CCD, respectively.

This is the final step in the pipeline.  The result is a science-quality spectrum extending from the blue limit of the blue camera to the red limit of the red camera.  Note that the wavelength pixel scale and the resolution will be discontinuous at the junction point because no resampling is performed.  It is written to a text file that also includes columns for the sky spectrum, an estimate of the variance spectrum, the $x$ pixel and $y$ profile centroids, the sensitivity function, and the fraction of constitutent image pixels flagged by the cosmic ray rejection algorithm.  This file includes extensive header meta-data describing all aspects of the instrument configuration and pipeline steps and settings.

\section{Additional features}

\subsection{Imaging pipeline}
\label{sec:imaging}

An imaging pipeline is also included.  Its early steps basically parallel the spectroscopic pipeline (overscan subtraction, flat-fielding, cosmic ray correction) using imaging-optimized routines.  From that point the behavior diverges completely as it performs astrometric and photometric registration using catalogs.  The ability to solve for and apply airmass solutions via spectrophotometric standards is also included---although this feature is rarely needed now after the release of the Pan-STARRS 3pi catalog, which covers almost any field visible from Keck.  (It is still important for $U$-band imaging.)  Distortion corrections to the astrometry have been investigated but are not generally performed, so the pipeline is not yet suitable for stacking imaging with large dithers (mosaicing).   A notable optional feature is the ability to create super-sky correction flats using specific blocks of imaging to remove dust features that have been observed to change positions as a result of changes to the red-side grating.    Unlike the spectroscopic pipeline (which is entirely written in IDL), the imaging pipeline has some important external dependencies - notably it requires SExtractor, SWarp, and a custom astrometric routine written in Python 2.4.  The imaging pipeline will be described in detail in future work.

\subsection{Notes about historical LRIS data}

LRIS-B has employed two detectors over its lifetime: an engineering-grade SITe 2048x2048 CCD that was used for a brief period after commissioning in 2000, and the science-grade CCD that was installed in June 2002 and remains in use.  Reductions of observations from the SITe detector are not yet supported.  The properties of the current instrument are quite stable, aside from occasional maintenance interventions that have changed the (otherwise fixed) wavelength range of each grism slightly (see \S \ref{sec:wavcal}).  The pipeline should be able to reduce LRIS-B data from any point in its history after the modern detector was installed.

LRIS-R has been equipped with three red detectors in its lifetime.  The most recent detector has been in use since December 2010 and has been stable over this period.  The two earlier detectors require some discussion.

The first red detector on LRIS was a standard thin CCD, employed from commissioning in the early 1990s until May 2009.  This detector differs from the current detector in many ways, including size (2048$\times$2048), pixel scale (0.210 instead of 0.135), quantum efficiency (much worse in the near-IR), and cosmic ray sensitivity (lower, i.e. better).  It is a single CCD with two amplifiers (rather than two CCDs with four amplifiers total) so there is no chip gap and no need to `split' the data.  It suffers from strong fringing in the near-IR, removal of which is extremely challenging.  (On-sky halogen flats and/or nodded pair subtraction are required for spectroscopic fringe correction; fringe frame subtraction is required for long-wave imaging.)  The pipeline supports LRIS R-1 data, but no fringe correction is applied, and the capabilities have not been tested on most gratings/configurations for that CCD.  Caution should be exercised in using the pipeline to process LRIS-red data prior to mid-2009.  

Very early observations with the red detector (2003 and earlier) lack some header metadata normally relied upon by by the pipeline, such as keywords about the dome orientation.  Different calibration lamps were also in use (the Cd and Zn lamps had not been installed, but Kr and Xe lamps were available).  These observations are unlikely to be successfully processed by the pipeline at all, although we anticipate adding support for LRIS data dating to this period eventually.

The second red detector was equivalent to the current one in most respects (size, pixel scale, layout, wavelength-dependent sensitivity, etc.).  It exhibited some curious noise and sensitivity patterns, including a significant noise enhancement (``tape noise'') close to the chip gap that rendered data close to the gap almost unusable\footnote{The third detector also exhibits this noise, but at a much reduced level.}.  A much bigger note of concern is that after a few months of installation (beginning approximately September 2009) this CCD began exhibiting CTE problems that are clearly visible in the form of one-directional trailing out of cosmic ray hits and sky lines.  Initially these affected only some amplifiers but soon appeared throughout the detector, and the issue was solved only by complete replacement of the detector in November 2010.  The pipeline will process data from this period as the CCD format is the same as the current detector, but there is no attempt made to correct for the CTE issues.  Results should be interpreted with caution.

\subsection{Interruption}

The pipeline is fully interruptible.  The user can Ctrl+C out of the reductions at any point and effectively resume the pipeline by repeating the same command it was started with.  The pipeline does not keep track of where it left off internally, but it can recognize when a step has already been performed based on the pre-existence of the output file it is planning to write---in these cases it will skip over that step.  To instruct the pipeline to skip over steps entirely, the user can state which stage in the processing sequence to begin (and, if desired, end) with.

Frequently it is desired to make changes and repeat steps (in particular with regard to aperture placement, see above).  By default, if apertures are changed and the pipeline is rerun, nothing will happen because this would require overwriting files generated by previous run.  However, a simple command-line option allows this behavior to be overridden, allowing a clean reduction (for an individual target or for the entire night) without having to selectively find and delete files.  (Re-)processing can also be restricted to specific steps or specific configurations.  The pipeline manual contains instructions and examples for this type of operation.

\begin{figure}[tbp]
\includegraphics[width=\textwidth]{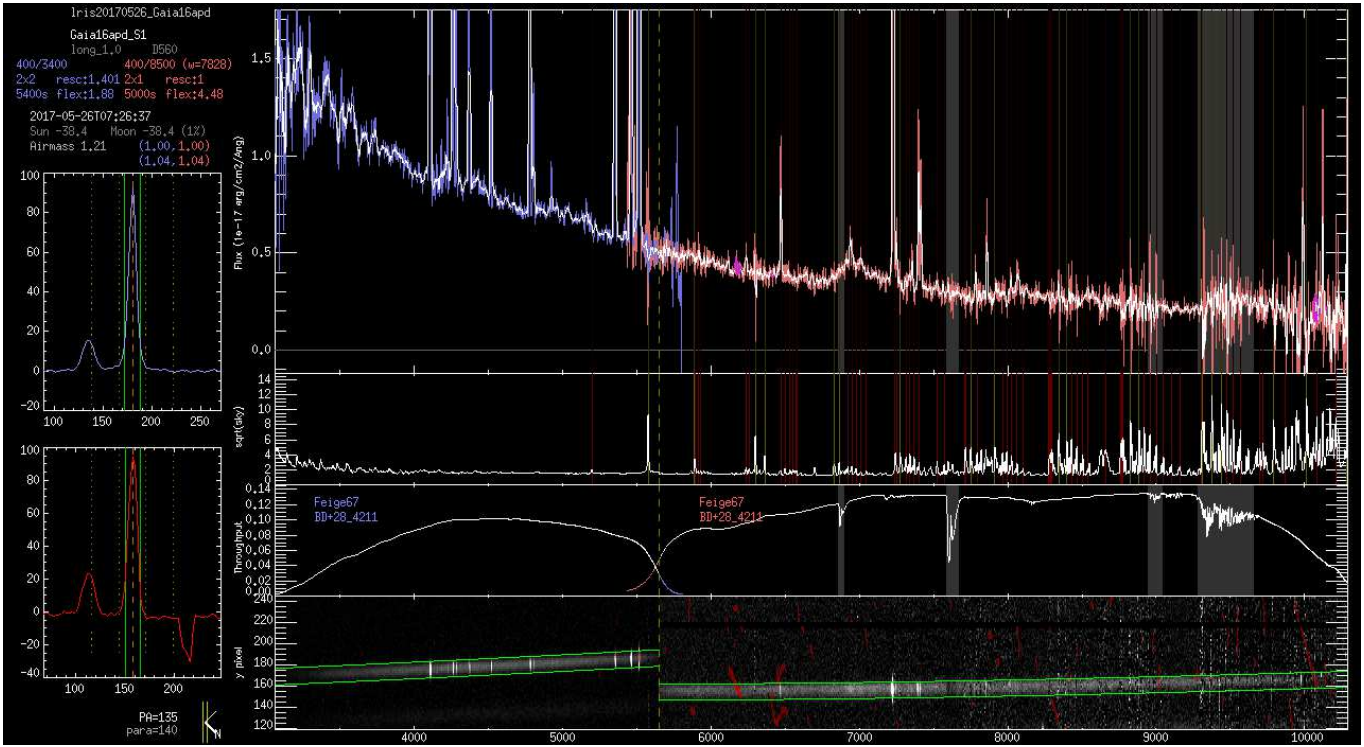}
\caption{Screenshot of the validation tool, showing spectra of an object and the sky, along with details of the extraction and flux calibration (throughput curve).  Telluric absorption bands, sky lines, and excised cosmic rays are also marked for reference.}
\label{fig:validation}
\end{figure}

\subsection{Identification of bad data}

The pipeline can recognize `bad' data in a variety of circumstances.  Flats with the lamp off or very low counts, observations with the shutter closed, saturated dome or twilight flats, saturated standards, and similar mistakes are identified via either the header keywords or counts in the image and once identified the pipeline will not conduct further processing.  The MIRA files used for segment alignment are also recognized automatically and not processed.  There is no need to formally delete these frames.  Indeed, it may be helpful to retain them to be sure the pipeline is able to faithfully record when instrument components were moved.

A frequent, minor observer error at the telescope is to forget to change the object name (written to the FITS headers under the OBJECT keyword) after slewing to a new source.  The pipeline does not depend on the OBJECT name in any formal sense: instead it uses TARGNAME which comes from the starlist uploaded at the beginning of the night.  (It does look for words like `sky' and `flat' to help distinguish twilight sky imaging frames in ambiguous cases but they are not required in any sense.)  However, if the telescope position and target are changed with no change in OBJECT name, the OBJECT name of subsequent entries will be set to the TARGNAME.

\subsection{Logging and diagnostics}

The pipeline reports what it is doing to the terminal as it proceeds.  Almost all of this information is also written to a log file.  This includes warning messages, or error messages if an important file cannot be processed for an unexpected reason such as a wavelength solution failure.  

A large number of check plots and summary plots are also written that document the wavelength solution and flexure correction, response curves, flux calibration (including telluric correction), tracing, and so on.  DS9 region files containing the output apertures are also written to disk and can be loaded to permit interactive user exploration of the tracing region.

Perhaps most usefully for the end scientific user, the pipeline also includes a validation tool that loads and plots an individual spectrum along with its corresponding sky spectra (for wavelength checking), thoroughput curves (for flux-calibration checking), and 2D spectra with overplotted trace curves as well as collapsed spatial profiles.  Pixels affected by cosmic rays or bad columns are marked.  With this tool the user can browse through spectra and identify any problematic results before they are ingested into online databases or used for scientific analysis.  The tool is demonstrated in Figure \ref{fig:validation}.

\section{Performance}

The processing time is dominated by the cosmic ray and sky-subtraction steps.  Some example run times for two recent observing nights are shown in Table \ref{tab:proctime}.  The pipeline is much slower when processing unbinned data compared to binned data.

\begin{table}
  \begin{center}
    \caption{Example processing times}
    \label{tab:proctime}
    \begin{tabular}{|l|l|l|}
      \hline
                    & Run 1 & Run 2\\
      \hline
      Date          & 2016-09-09 &  2017-07-15 \\
      $N_{\rm cal}$ &   18 & 119     \\
      $N_{\rm sci}$ &   66 &  56     \\
      binning       &  1x1 & 2x1/2x2 \\
      \hline
      \textbf{Step} & \multicolumn{2}{c|}{\bf Time (s)} \\
      \hline
      prepare     &  92 &  86 \\
      makeflat    &  17 &   9 \\
      flatten     &  26 &   9 \\
      split       &   4 &   2 \\
      crclean     & 595 & 217 \\
      skysubtract & 557 & 175 \\
      sum         &  41 &  19 \\
      trace       &  20 &   7 \\
      extract     &  43 &  18 \\
      wavcal      &   3 &   3 \\
      wavapply    &  25 &  26 \\
      response    &   5 &   2 \\
      fluxcal     &   7 &   6 \\
      connect     &   5 &   4 \\
      \hline
      total       &1438 & 583 \\
      \hline
    \end{tabular}
  \end{center}
\end{table}

Reducing imaging data is substantially slower (nights primarily or entirely taken in imaging mode can require over an hour to reduce), mostly because of the larger number of files involved (shorter exposures).

Data products are usually science-quality, as long as observations satisfy the basic requirements and the aperture background regions are reasonably ``clean'' (e.g. free of host-galaxy flux).  They are not necessarily the best reductions possible: use of optimal extraction could improve S/N significantly, and the flux-calibration accuracy is typically limited by how clean the spectrum of the flux-standard was or how much red-blue overlap is provided by the setup.  Given the difficulty of accurately flux-calibrating (and occasionally tracing and/or wavelength-calibrating) over the dichroic overlap region at the CCD edges it is not uncommon to observe small jumps or discontinuities in this area in final spectra, and users should be aware that there is always intrinsic uncertainty in the red-blue scaling for faint objects.  Other non-trivial problems are much rarer, although subtle errors can occasionally be introduced at all stages (see \S \ref{sec:archive}).

Most apparent failures in processing are associated with missing calibration files (or equivalently, cases where all calibration files acquired in a given setup are identifiably bad---and therefore flagged and not processed by the piepline).  Such issues can of course be solved by acquiring/repeating the relevant calibrations: at the telescope (if the pipeline is being employed during real-time reductions of an observing run), or by searching for an appropriate set of calibration files in the public archive and placing them in a pipeline-accessible reduction directory.  Future pipeline releases will include a set of archival calibrations for standard observing modes, as well as include an option to continue with no flat-fielding.

\begin{figure}[tbp] 
   \centering
   \includegraphics[width=7.1in,angle=0]{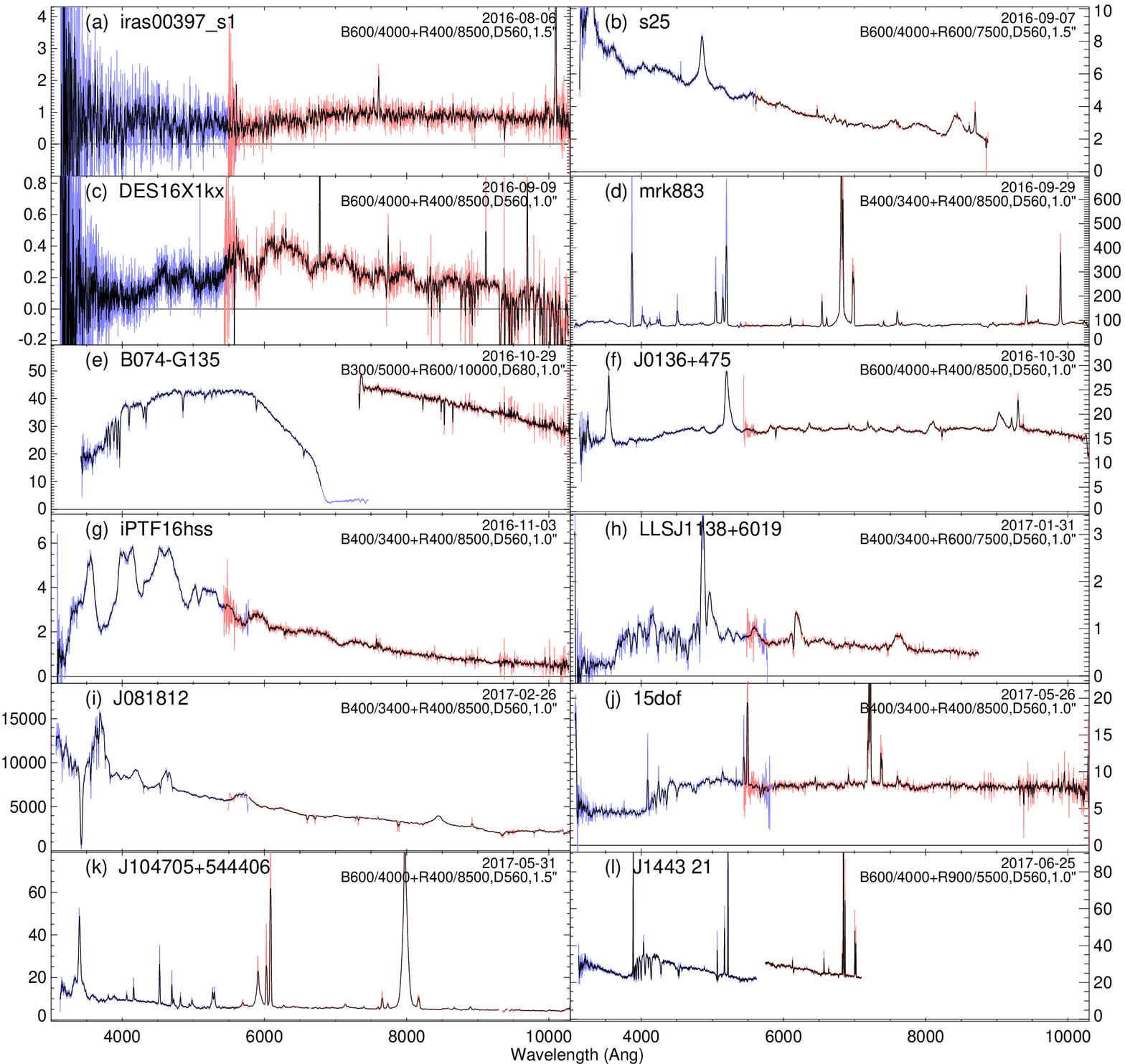}
   \caption{\small
   Examples of data processed by the pipeline (flux units are 10$^{-17}$ erg cm$^{-2}$s$^{-1}$\AA$^{-2}$).  Red and blue curves show the separate (rescaled) blue and red spectra;  the black curve shows the joint spectrum convolved with a smoothing function.  Most show no significant problems.  Sky lines are oversubtracted from spectrum (c), probably due to imperfect flatfielding.  Spectrum (e) was taken with the 300/5000 grism, which is contaminated by second-order light. Spectrum (l) has no red/blue overlap, so there is a rescaling jump.  
   Associated programs: C243LA (Harrison), C258LA (Djorgovski), U150LA (Filippenko), C204LA (Cohen), Y053M (van Dokkum), C241LA (Ravi), C265LA (Prince), U147LA (White), W224 (Cooke), C299 (Kulkarni), C275 (Djorgovski), U085 (Canalizo).
 }
\label{fig:examples} 
\end{figure}

\section{LRIS Archive Reduction}
\label{sec:archive}

To demonstrate the efficacy of the pipeline in practice and in the interest of making scientific data accessible to the worldwide community, we downloaded all LRIS observations from the two most recent publicly-released semesters in the Keck Observatory Archive (2016B and 2017A, spanning from August 2016 to July 2017).  The data were sorted by observing night, and the pipeline was scripted to run in fully automated mode in each nightly directory.

To allow for the possibility that flats might be obtained on nights different from the science observations, we instructed the pipeline to search the reduction directories from other nights if flat-field files or arc-calibration files needed for science calibration could not be found in the nightly directory.  No other modifications were made, and we did not supervise the reductions while they were in progress and did not inspect or delete any input files, relying entirely on the pipeline to remove irrelevant or problematic calibrations and choose the most suitable calibration files to use for processing science data.  Results were examined (using the validation tool) only following the completion of all processing.  (Some examples of successful final spectra are shown in Figure \ref{fig:examples}.)

The pipeline successfully processed (i.e., used in producing final output spectra) 2234 out of 2549 science exposures (88\%).  In terms of exposure time, the fraction is 1532012 out of 1696211 seconds (90\%)\footnote{Both these statistics include combined numbers for red and blue cameras.   We define a science exposure as an exposure not of a known standard star, of at least 45 seconds duration, and in 12-degree twilight or darker skies.}.  We did not carefully inspect all examples of processing failures to ascertain the specific cause, although the large majority appeared to be due either to the lack of detectable continuum to trace/extract (often as the result of cloudy weather or a very diffuse target such as a nearby galaxy), to a lack of usable arc calibrations with fully warmed lamps, or to a lack of a usable standard star observed on the same night.  Observations in the latter two categories should be recoverable in future releases by use of a standard reference arc and response curve solutions.

Most of the output spectra were judged to be scientifically usable in validation: out of 704 combined ``final'' spectra, 491 were judged to have no issues or only minor issues (e.g. issues at the edges of the spectra, small rescaling jumps, small telluric residuals, etc.).  We identified 184 spectra to have more significant issues (i.e., at a level for which a typical scientific user would need to correct before performing scientific analysis or publishing), but these would either be correctible without re-reduction or they affected only a small part of the spectrum.  Only 29 were judged to be mostly or entirely unusable.  

Problems had a diverse array of causes.  For spectra with ``significant'' but non-fatal issues, the most common problem was  flux calibration.  Most of these involved data taken with the 300/5000 blue grism (which cannot be calibrated accurately across the full wavelength region, and essentially every observation taken with this setup was flagged as problematic).  On two nights clearly problematic standard-star exposures (excessive flux over the dichroic) were seen in other grisms.  These likely originated due to failure of the blue shutter at the end of the exposure (a common hardware failure mode on LRIS).  Significant telluric calibration residuals (producing false emission lines, etc.) were also a relatively common issue---most commonly as a result of the cosmic ray cleaner over-cleaning the narrow series of absorption lines covering the NIR telluric water vapor bands in the standard-star exposure.  Poor sky subtraction of the NIR OH lines was seen in a number of faint-object spectra (generally on nights for which halogen flat lamps were used, although the most severe cases were associated with a run in which the red trace position fell unusually close to the chip gap where tape noise compromises the accuracy of dome flats as well).  Issues with tracing, rescaling, cosmic-ray false-positives, or aperture background contamination were all also observed but generally rarer, cumulatively affecting only a few percent of all input spectra.

The most common issue for failed spectra\footnote{These statistics are for spectra for which final reductions were produced but were rejected at the validation stage.  They do not include spectra which could not be extracted or calibrated at all (about 10\% of all spectra, see the discussion earlier in this section).} (affecting 10 out of 26) was simply that the pipeline extracted a ``source'' where no source was actually apparent on the trace (a situation for which the associated data would probably be unusuable anyway).  Six spectra were also rejected due to severe flux calibration issues, almost all of which were associated with a night in which a broad-band imaging filter was stuck within the optical path of the red camera for one standard-star observation.  Five were rejected due to background contamination issues and four due to a failure of the red-blue alignment procedure (resulting in ``chimeric'' spectra hybridizing two unrelated objects.)

Most of the above issues are fixable: a small amount of user oversight would be able to, for example, recognize and exclude problematic standards or re-process certain observations with a lower cosmic-ray rejection threshold.  Many issues could also be recognized and fixed automatically by the pipeline itself, and these will be explored in future releases.

All spectra, including ``failed'' spectra, have been made available online\footnote{The provisional location is: \url{http://www.astro.ljmu.ac.uk/~aridperl/lris/archive/}}.  In the near future we will expand our reductions back at least as far as the commissioning date of the upgraded red detector (2009), and attempt to salvage observations that are not currently optimally processed in unsupervised mode.  These spectra will be made permanently available as a user-provided data set at the official Keck archive\footnote{\url{https://koa.ipac.caltech.edu/}}.

Users interested in making use of these spectra for scientific analysis should be mindful of the limitations discussed earlier.  In particular, the S/N will be lower than what would be available from an optimum extraction, there may be issues in extraction of faint targets at the far blue/red ends or in flux calibration across the dichroic junction, and the wavelength calibration may be short of its theoretical optimum and may (particularly on the blue side) not always be flexure-corrected.  On rare occasions bright emission lines may be wrongly removed as cosmic rays.  Observing runs in which no high-quality standards were observed may also show flux calibration deviations of up to 20\% due to standard-star line residuals (especially if reference spectra for the relevant standards are poor), and high-airmass objects will show telluric residuals if standards were not obtained at similar airmass and seeing conditions.  Users should also beware that LRIS does not have any straightforward method of robustly tracking slit placement (e.g., slit-viewing images): there is no guarantee that the target specified by the user (and given as the TARGNAME or OBJECT header value) was accurately placed on the slit or extracted by the pipeline.

Users are advised to download the latest version of the pipeline and re-reduce any relevant observations (potentially using different settings if, e.g., accurate extraction of extended sources is needed, if cosmic rays are being over- or under-corrected, if a flux standard is problematic, etc.).  Almost all LRIS data prior to the standard 18-month proprietary period is available online and a tool exists in the pipeline toolkit to queue direct downloads of raw data from the Keck archive.

\section{Conclusions}

In this paper we have demonstrated a robust, effective, efficient pipeline for fully automated processing of LRIS observations.  The pipeline is generic to all configurations of the longslit and requires no user input, except to control fine details of aperture extraction.  While highly customized for managing the many complexities of LRIS (flexure, separate red+blue observations, etc.), the design of the pipeline is generic and could easily be extended to other facilities.

The pipeline is currently available at \url{http://www.astro.caltech.edu/~dperley/programs/lpipe.html} and will also be made available on github and other package repositories in the near future.  

\acknowledgments

This research has made use of the Keck Observatory Archive (KOA), which is operated by the W. M. Keck Observatory and the NASA Exoplanet Science Institute (NExScI), under contract with the National Aeronautics and Space Administration.  DAP would like to thank the observatory and the archive for their support of the publication of this paper, as well as current and former LRIS support astronomers (in particular, Greg Wirth, Marc Kassis, and Luca Ricci) for their advice and assistance during many previous observing runs.
Additional thanks are provided to the referee (Dan Kelson) for constructive and helpful comments on the manuscript and future suggestions for improvements to the pipeline, and to pipeline users (including Tom Brink and Kevin Burdge) for additional feature suggeestions.  The author also wishes to recognize and acknowledge the very significant cultural role and reverence that the summit of Maunakea has always had within the indigenous Hawaiian community.  We are most fortunate to have the opportunity to conduct observations from this mountain. 

\facility{Keck:I (LRIS)} 
 


\end{document}